\documentstyle[11pt,newpasp,twoside,epsf]{article}
\def\edcomment#1{\iffalse\marginpar{\raggedright\sl#1\/}\else\relax\fi}
\reversemarginpar

\begin{document}
\title{X-ray temperature and morphology of z$>$0.8 clusters of galaxies}
 \author{B. Holden}
\affil{University of California, Davis \& Insitute for Geophysics and
 Planetary Physics, L-413, P.O. 808, Livermore, CA 94551-0808}
\author{S. A. Stanford}
\affil{University of California, Davis \& Insitute for Geophysics and
 Planetary Physics, L-413, P.O. 808, Livermore, CA 94551-0808}

\author{P. Rosati}
\affil{European Southern Observatory, Karl-Schwarzschild-Str. 2,
D-85748 Garching bei M\"{u}nchen}

\author{P. Tozzi}
\affil{Osservatorio Astronomico di Trieste,Via G. B. Tiepolo, 11,
34131 Trieste} 

\author{G. Squires}
\affil{California Institute of Technology, MS 105-24, 1201 East
California Blvd, Pasadena, CA, 91125}

\author{S. Borgani}
\affil{INFN, Sezione di Trieste, Via A. Pascoli, I-06100 Perugia}

\author{P. Eisenhardt}
\affil{Jet Propulsion Laboratory, California Institute of Technology,
MS 169-327, 4800 Oak Grove Drive, Pasadena, CA, 91109}

\begin{abstract}

We discuss our current progress in studying a sample of $z>0.8$
clusters of galaxies from the {\tt ROSAT} Distant Cluster Survey.  To
date, we have Chandra observations for four of the ten clusters.  We
find that the morphology of two of these four are quite regular, with
deviations from circular of less than 5\%, while two are strikingly
elliptical.  When the temperatures and luminosities of our sample are
grouped with six other high-redshift measurements, there is no
measured evolution in the luminosity-temperature relation.  We
identify a number of X-ray emitting point sources that are potential
cluster members.  These could be sources of intracluster medium
heating, adding the entropy necessary to explain the cluster
luminosity-temperature relation.

\end{abstract}

\section{Introduction}

Our sample of $z>0.8$ clusters of galaxies is part of the {\tt ROSAT}
Distant Cluster Survey (RDCS; Rosati {\em et al.} 1998).  The RDCS
contains 137 clusters of galaxies covering 50 deg$^2$ to an X-ray flux
limit of $1 \times 10^{-14} {\rm erg\ s^{-1}\ cm^{-2}\ }$ in the
0.5-2.0 keV band.  This includes the highest redshift, X-ray
selected clusters galaxies known to date, with ten clusters at $0.8 \le
z \le 1.3$.  The luminosity and redshift distribution of our high
redshift sample is plotted in Figure 1, along with that distribution
for the Bright SHARC sample (Romer {\em et al.}  2000) and the
Einstein Medium Sensitivity Sample of clusters of galaxies (Henry {\em
et al.} 1992; Gioia \& Luppino 1995) for comparison.

We have targeted this sample for follow-up with the two premier X-ray
observatories, Chandra and XMM-Newton.  To date, we have Chandra
observations for four clusters in that sample, with the details listed
in Table 1.  The X-ray data for two of the clusters discussed here, RX
J0848+4453 and RX J0849+4452, are presented in Stanford {\em et al.}
2001.  For the rest of this work we will assume $\Omega_m = 0.3$,
$\Omega_{\Lambda} = 0.7$ and ${\rm H_{\circ}} = 65\ {\rm km\ s^{-1}\
Mpc^{-1}}$.

\begin{table}
\caption{Summary of Chandra Observations}
\begin{tabular}{l|ccccc} 
\tableline
Name & $\alpha$ & $\delta$ &  z &  Lum. & Temp. \\
     & (J2000)  & (J2000)  &    &  ($10^{44}\ {\rm erg\ s^{-1}}$) & (keV) \\
\tableline
RX J0848+4453 & 08 48 35.8 & +44 53 45.5 & 1.26 & $0.64^{+0.25}_{-0.16}$ & $1.6^{+0.8}_{-0.6}$ \\[3pt]  
RX J0849+4452 & 08 48 58.7 & +44 51 53.3 & 1.27 & $3.3^{+0.7}_{-0.5}$ & $5.8^{+2.6}_{-1.7}$ \\[3pt]  
RX J0910+5429 & 09 10 44.9 & +54 22 07.7 & 1.10 & $2.0^{+.3}_{-0.2}$ & $7.2^{+2.2}_{-1.4}$ \\[3pt]  
RX J1317+2911 & 13 17 21.7 & +29 11 18.1 & 0.80 & $1.8^{+0.7}_{-0.4}$ & $3.7^{+1.5}_{-0.8}$ \\[3pt]  
\tableline
\tableline
\end{tabular}
\end{table}

\begin{figure}
\begin{center}
{\centering \leavevmode
\epsfxsize=0.8\textwidth \epsfbox{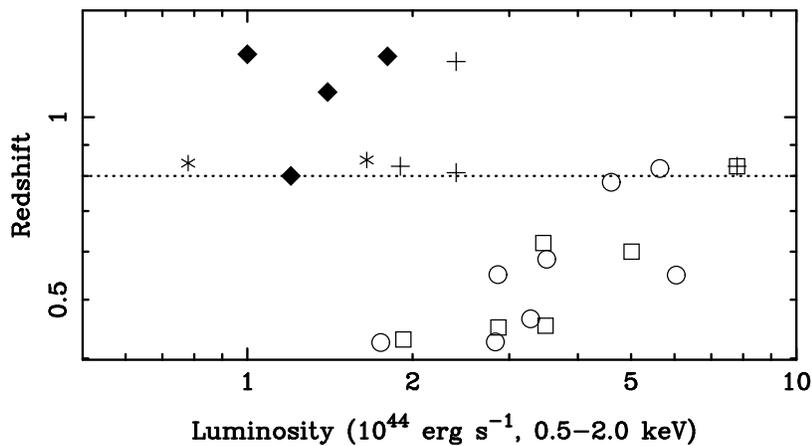}}
\end{center}
\caption{
The luminosity and redshift distribution of our sample
compared with the Bright SHARC (open squares) and the EMSS (open
circles).  The four  diamonds are discuss in this work, while the
remaining are either pending observations (plus symbols) or not
observed (asterisks).  The dotted line shows our redshift limit.
The values of the luminosities are the catalog values as measured by
the RDCS, not values from the new Chandra data
}
\end{figure}

\section{Morphology}

\begin{figure}[ht]
\begin{center}
{\centering \leavevmode
\epsfxsize=0.7\textwidth \epsfbox{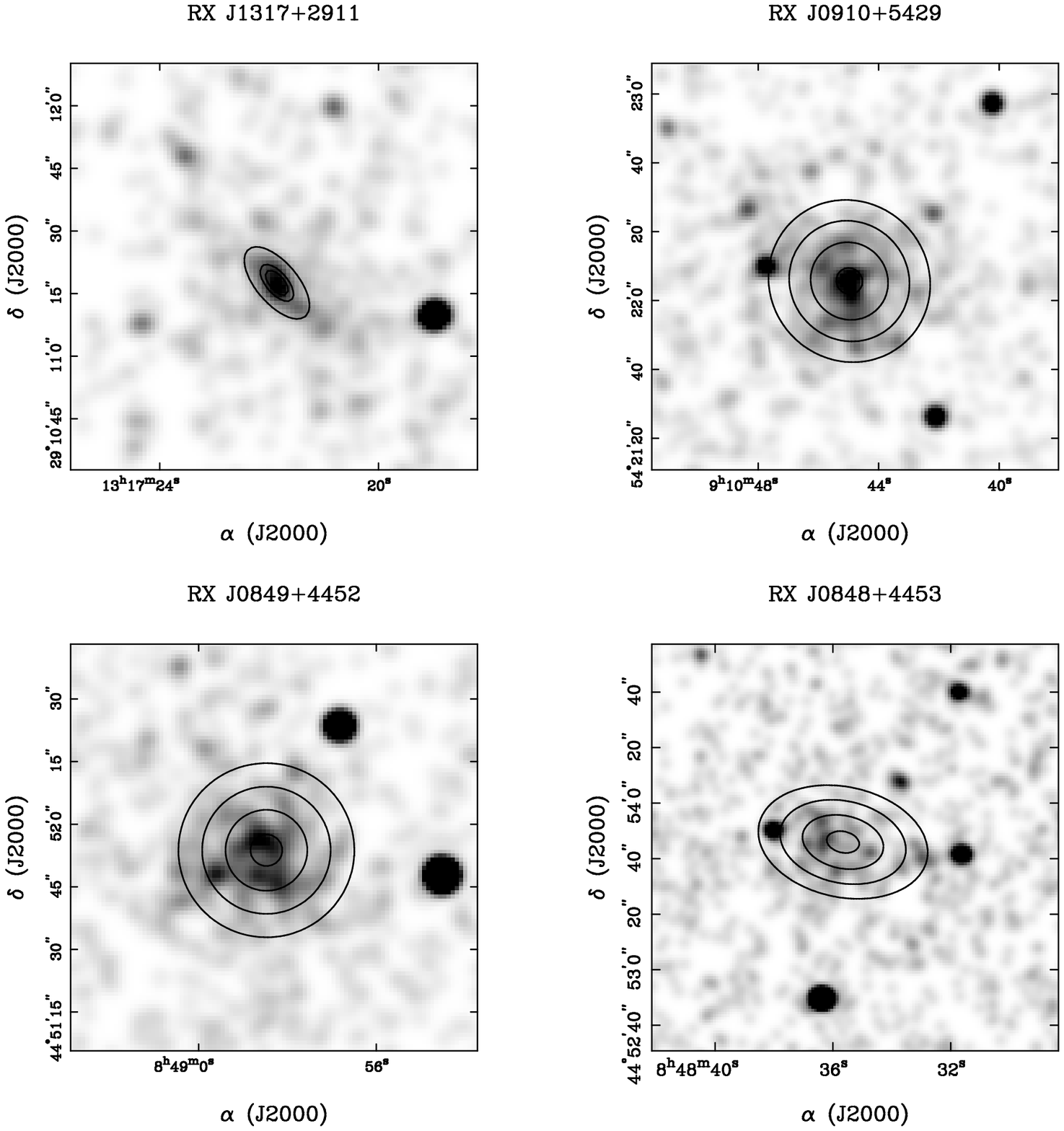}}
\end{center}
\caption{ Images of the four clusters in our sample.  We plot the
smoothed photon distribution with the best fitting $\beta$ model overlaid as
contours.  Each model has four contour levels corresponding to 90\% of
the peak value, 50\% of the peak value or the core radius, 25\% of the
peak value and 12.5\% of the peak value.  The exception is RX
J1317+2911 which has contours of 12.5\%, 3\%, 2\% and 1\% of the peak
amplitude.  }
\end{figure}

Quantitative morphology for clusters of galaxies usually means fitting
a $\beta$ model.  We fit two-dimensional models to the 0.5-2.0 keV
photon distributions for the four clusters of galaxies in our sample,
after excluding point sources.  The model included a constant term for
the background in addition to the cluster model.  These results are
shown in Figure 2 and in Table 2.

\begin{table}
\begin{center}
\begin{tabular}{l|cc}
\tableline
Name & $\beta$ & ${\rm r_c}$ \\
     &         &  (kpc)      \\
\tableline
RX J1317+2911 & 0.3 & 1 \\[3pt]  
RX J0910+5429 & $0.89^{+0.24}_{-0.23}$ & $171^{+49}_{-57}$ \\[3pt]   
RX J0849+4452 & $0.73^{+0.31}_{-0.15}$ & $81^{+23}_{-23}$  \\[3pt]    
RX J0848+4452 & 0.8 & 200 \\[3pt]  
\tableline
\tableline
\end{tabular}
\end{center}
\end{table}

Two things are striking.  First, two of the models are quite round.
The deviations from a circular model are at less than 5\% for both.
This is in contrast with, say, MS 1054-0321 or RX J0152.7-1357, which
show greatly disturbed, asymmetric morphologies.  Second, for these
round clusters, the best fitting core radii and values of $\beta$ are
normal for low redshift clusters of galaxies.  These two clusters are
also have the largest number of net counts.  For RX J0910+5429, in the
0.5-2.0 keV band we measured a total of 500 counts, including the
background, within two core radii. We expected 137.4 counts from the
background in the same bandpass. With RX J0849+4452, we found 469
counts in the 0.5-2.0 keV band within two core radii while expecting
142.8 from the background.  In contrast, RX J0848+4453 has only 259
counts with 104.6 background counts expected and RX J1317+2911 has 217
events with 54.7 expected.  Therefore, we may only be resolving the
cores of these systems, and missing a smooth outer region because of
the lack of events.  Nonetheless, both have strange morphologies not
well described with a $\beta$ model.  Because of the poorer statistics
and the strange morphologies for these clusters, we quote no error
bars on our measured parameters for the $\beta$ model.  RX J0848+4453
is also strongly contaminated by point sources, see Figure 2.  Two of
these point sources are potential cluster members, based on
photometric redshifts, and are likely to be active galactic nuclei
based on their hardness ratios.  After excluding these sources, the
flux we measured for RX J0848+4453 was significantly lower than what
we measured in the original RDCS.

For the two clusters with well constrained morphologies, RX J0849+4452
and RX J0910+5429, we used an  aperture of twice the core
radius to measure the temperature and flux.  We than use the best
fitting $\beta$ model to compute the total flux.  In Table 1,
we quote the best fitting temperatures and luminosities.  For the two
remaining clusters, we used a curve of growth analysis to pick a total
flux aperture.  We used that aperture to measure the temperature as
well.  These results are also included in Table 1.

\section{Luminosity-Temperature Relation}

We plot, in Figure 3, the luminosities and temperatures for our data
and a number of other high redshift clusters.  For comparison, we also
plot the relation for low redshift clusters of galaxies as measured by
Markevitch (1998) and the low redshift group data of Helsdon \& Ponman
(2000).  When our data are combined with the other high redshift clusters
of galaxies, it appears that there is minimal or no evolution in the
luminosity-temperature relation.

\begin{figure}[h]
\begin{center}
{\centering \leavevmode
\epsfxsize=0.7\textwidth \epsfbox{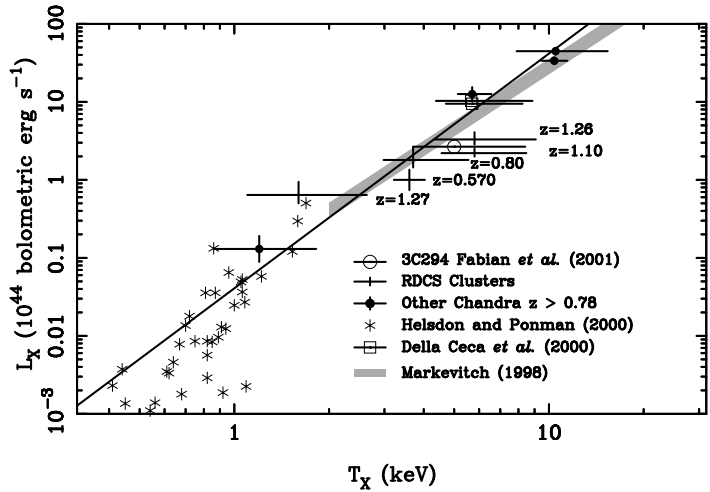}}
\end{center}
\caption{The Luminosity-Temperature relation for high redshift
clusters of galaxies.  The ${\rm z} > 0.78$ clusters come from Borgani
{\em et al.} (2001).  The measurements from della Ceca {\em et al.}
(2000) are only those clusters with ${\rm z} > 0.8$, not the entire
sample in that paper.  The points marked by redshifts are various RDCS
clusters, including the cluster at z=0.570 from Holden {\em et al.}
(2001).  The solid line represents are best fit to the relation.}
\end{figure}

As a rough test for evolution, we fit the relation $L_{Bol} \propto
T^{\alpha}$ to the high redshift data, including the four clusters in
our sample, the two $z>0.8$ clusters (RX J1716.6+6708, and RX
J0152.7-1357) from della Ceca et al. (2000), three clusters
(MS1137.5+6625, 1WGAJ1226.9, \& CDFS-CL1) summarized in Borgani et
al. (2001) and the results for MS1054.4-0321 from Jeltema {\em et al.}
(2001) (see also these proceedings).  We fit the relation using the
method of Akritas and Bershady (1996) which accounts for errors in
both the temperature and luminosity.  For our sample of ten clusters
and groups, we found the best fit slope to be $\alpha = 3.0 \pm 0.5$
(90\% confidence limits) at our median redshift of $z = 0.83$.  Our
measured slope differs by slightly more than one standard deviation from
$\alpha = 2.63 \pm 0.27$ (90\% confidence limits), the relation of
Markevitch (1998).  Our result is in good agreement with the slope of
$\alpha = 3.1 \pm 0.6$ from Allen \& Fabian (1998).  We note here that
we fit only isothermal models to all of the clusters in our sample.
Therefore, we compare our results with the Model A from Allen \&
Fabian (1998), which used similar assumptions.

\section{A Potential Source of Intracluster Heating?}

\begin{figure}
\plotone{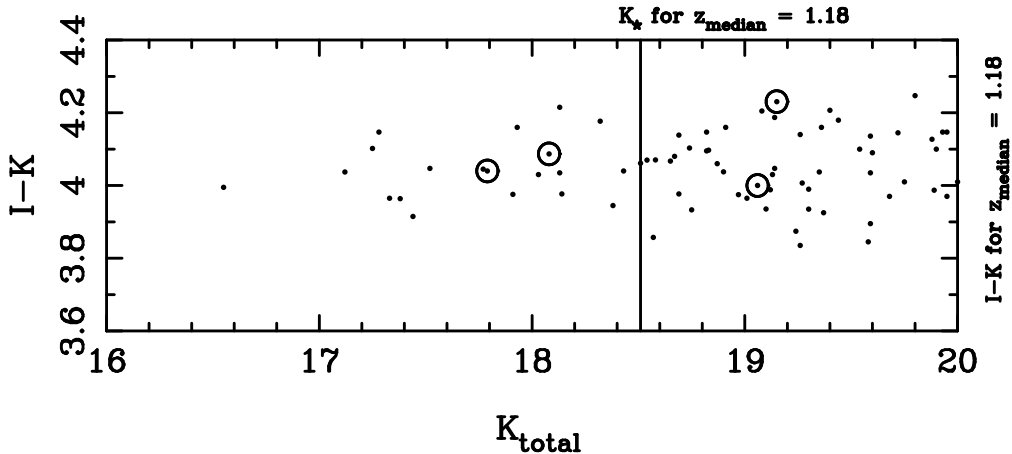}
\caption{ The color-magnitude diagram for all four clusters in our
sample.  The colors and magnitude for each cluster galaxy shifted to
the median redshift of our sample, $z=1.18$.  Circled objects are
galaxies within one standard deviation of the color-magnitude relation
and are X-ray point sources less than 2\arcmin\ from the cluster
center.  }
\end{figure}

In Figure 4, we plot a color magnitude diagram for the 4 clusters in
our sample.  We have shifted the colors and magnitudes to the median
redshift for our sample of 1.18.  We circle in X-ray point sources
that are within one standard deviation of the mean color-magnitude
relation and within 2\arcmin\ of the cluster center.  So, each of
these galaxies are possible cluster members and they appear to have
X-ray emission.  The number of these objects in entirely consistent
with the results of Barger et al. (2001) who find 4\% of $L_{\star}$
or bright galaxies are X-ray emitters and 7\% of their entire sample.
As the potential sources of the X-ray emission are active galactic
nuclei or a very hot interstellar medium, these objects could be
important sources of entropy for the intracluster medium (ICM), e.g.,
Bower (1997) or Ponman, Cannon \& Navarro (1999).  In Figure 5, we
show a hard X-ray image of RX J0910+5429 along with the same soft
image shown in Figure 2.  There is an obvious hard excesses to the
south of the core of the cluster.  The hard excess is statistically
significant at the 99\% confidence limit when compared with both the
central region of the cluster (within an equal area centered slightly
to the north of the centroid) and when compared with the background.
One explanation is we are observing a merger event and the resulting
shock.  Another is the circled X-ray source in the southeast of the
cluster, one of the candidates in Figure 4, is heating the ICM.  

\begin{figure}
\plotone{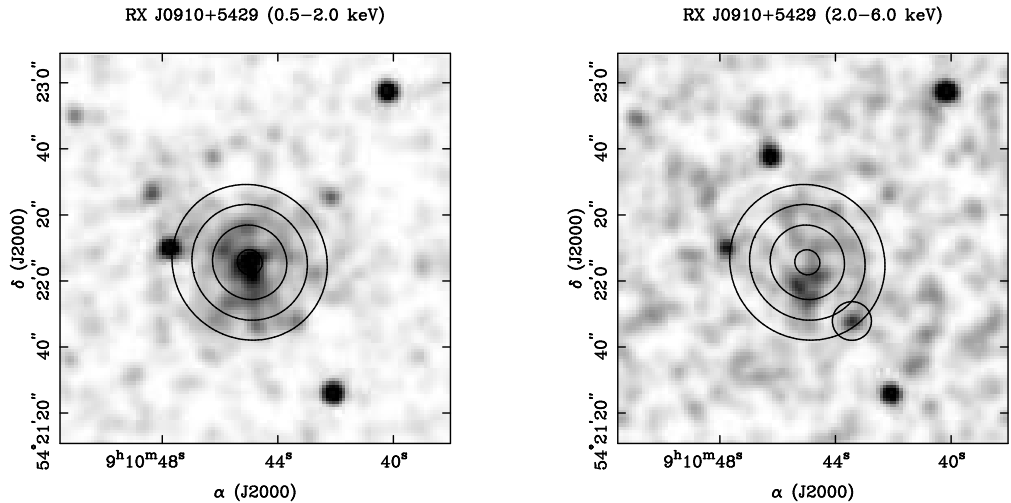}
\caption{ Images of RX J0910+5429 in the ROSAT band and in a hard band
from 2.0 - 6.0 keV.  We plot the smoothed photon distribution with the best
fitting $\beta$ model overlaid as contours.  Each model has four
contour levels corresponding to 90\% of the peak value, 50\% of the
peak value or the core radius, 25\% of the peak value and 12.5\% of
the peak value.  The circle, in the right panel, is around a potential
hard X-ray emitting cluster member.}
\end{figure}

Therefore, if we understand these hard X-ray sources and, most
importantly, learn how much energy and over what range redshifts they
are adding to the ICM, we can directly test the idea of AGN heating of
the ICM (see Valageas \& Silk 1999; Wu, Fabian \& Nulsen 2000)
and answer the question of the origin of the ICM pre-heating.

\section{Summary}

We have observed with the Chandra telescope four clusters of galaxies
in our sample of ten at $z>0.8$.  Two of these four are relaxed
looking clusters with values for $\beta$ and the core radius entirely
consistent with low redshift clusters.  The other two, however, have
elongated morphologies and, in one case, an extreme value for $\beta$
and the core radius.

Despite the wide range of morphologies, all of our clusters
agree with the low redshift luminosity temperature relation.
Including other results, we can see that the L-T relation has little
evolution over almost one order of magnitude in temperature out to a
$z_{median} = 0.83$.

Finally, we identify a potential source of cluster members for the
additional entropy needed to explain the L-T relation.  Further
investigation into these objects could shed light on the origin and
evolution of the intracluster medium.


\begin{references}

\reference Akritas, M. G. \& Bershady, M. A.  1996, \apj, 470, 706

\reference Allen, S. W. \& Fabian, A.  1998, \mnras, 297, L57

\reference Barger, A., Cowie, L. L., Mushotzky, R. F., \&  Richards,
E. A.  2001, \aj, 121, 662

\reference Borgani, S., Rosati, P., Tozzi, P., Stanford, S. A.,
Eisenhardt, P., Lidman, C., Holden, B., della Ceca, R., Norman, C., \&
Squires, G.  2001, \apj, submitted, astro-ph/0106428

\reference Bower, R. G.  1997, \mnras, 288, 355

\reference della Ceca, R., Scaramella, R., Gioia I.M., Rosati, P.,
Fiore, F. \& Squires, G.  2000, \aap, 353, 498

\reference Fabian, A., Crawford, C. S., Ettori, S. \& Sanders, J. S.
2001, \mnras, 322, L11  

\reference Gioia, I. \& Luppino, G.  1994, \apjs, 94, 583

\reference Helsdon, S. F. \& Ponman, T. J.  2000, \mnras, 315, 356 

\reference Henry, J. P., Gioia, I. M., Maccacaro, T., Morris, S. L.,
Stocke, J. T. \& Wolter, A.  1992, \apj, 386, 408

\reference Holden, B., Stanford, S. A., Rosati, P., Squires, G.,
Tozzi, P., Fosbury, R. A. E., Papovich, C. Eisenhardt, P., Elston, R.,
\& Spinrad, H. 2001, \aj, 122, 629

\reference Markevitch, M.  1998, \apj, 504, 27

\reference Ponman, T. J., Cannon, D. B., \& Navarro, J. F.  Nature,
397, 135

\reference Romer, A.K., Nichol R.C., Holden, B. P., Ulmer, M.P., Pildis, R.A.,
Adami, C., Burke, D.J., Collins, C.A., Merrelli, A.J., \& Metevier,
A.M.  2000, \apjs, 126, 209

\reference Rosati, P., Della Ceca, R., Norman, C., \& Giacconi, R.
1998, \apjl, 492, L21

\reference Stanford, S. A., Holden, B. P., Rosati, P., Tozzi, P., Borgani,
S., Eisenhardt, P., \& Spinrad, H.  2001, \apj, 552, 504

\reference Valageas, P. \& Silk, J. 1999 \aap, 347, 1

\reference Wu, K. K. S., Fabian, A. C., \& Nulsen, P. E. J.  2000,
MNRAS, 318, 889 

\end{references}
\end{document}